\documentclass{aa}  

\usepackage{graphicx}
\usepackage{txfonts}
\usepackage{lipsum}
\usepackage{subcaption}         
\usepackage{lscape}             
\usepackage{placeins}           
\usepackage[normalem]{ulem}

\usepackage[colorlinks=true, linkcolor=blue, citecolor=blue, urlcolor=blue]{hyperref}

\begin{document}

   \title{Infrared spectra of methane-containing ice mixtures for JWST data analyses}



   \author{V. Karteyeva
        \and R. Nakibov \and I. Petrashkevich \and M. Medvedev \and A. Vasyunin}

   \institute{Research Laboratory for Astrochemistry, Ural Federal University, Kuibysheva St. 48, Yekaterinburg 620026, Russia\\
             \email{varvara.karteeva@urfu.ru}
            }

   \date{Received }

 
  \abstract
   {Solid methane (CH$_4$) is an important molecule in interstellar and planetary environments, serving as a precursor to complex organic compounds, and it is a potential biosignature in exoplanetary studies. Despite its significance, laboratory data on the low-temperature phase of methane below 10 K remain limited.}
   {We obtained spectra of methane in binary mixtures at 10 K and compared them to the spectra obtained at 6.7 K. These temperatures correspond to phases II and II* of pure methane and are representative of dark molecular clouds and protostars in early stages. We also tested whether the data we obtained can be applied to interpret JWST data. }
   {Laboratory reference spectra were obtained with the ISEAge setup via Fourier transform infrared spectroscopy in transmission mode. A weighted~$\chi^2$ minimization was used for the fitting.}
   {We present infrared spectra with corresponding band strengths of pure methane and binary mixtures with methane: CH$_4$:H$_2$O, CH$_4$:CO$_2$, CH$_4$:CH$_3$OH, and CH$_4$:NH$_3$ at 6.7 K and 10 K. They show an increase of 20\% in mixtures compared to the commonly used 10 K band strength value of pure methane. We also tested whether the spectra can be used on open JWST data by probing the spatial distribution of methane in B335. We also present additional experiments concerning the phase transition of methane between phase II* and phase II.
   }
   {Our results reveal distinct spectral features for methane in non-H$_2$O environments that enable a more accurate interpretation of JWST observations. The dataset of spectra is publicly available on Zenodo and can be used for fitting JWST data.}


   \maketitle \nolinenumbers

\section{Introduction}
Solid methane is the most abundant and one of the simplest hydrocarbon molecules found in pre-stellar objects and protostars \citep{spitzer_oberg_2008}. Both theoretical  \citep{2008ApJ...681.1385H,2008ApJ_CH4_chains,2021MNRAS.508.1526C} and experimental \citep{2002A&A...384..343B,2009Icar..200..338H,2018PCCP...20.5435A,2018ApJS..234...15Z} studies have considered methane a precursor for more complex carbonaceous molecules. It is also considered a biosignature for exoplanetary studies \citep{2018AJ....156..114K,thompson2022case,2025AREPS..53..283G}. Methane is a part of the atmosphere of planets, for instance Jupiter and Mars \citep{ gautier1982c,2004Icar..172..537K}, and exoplanets, for example HD 209458b and HR 8799 b \citep{2009ApJ...704.1616S,barman2015simultaneous}.

The first discovery of interstellar solid methane was made with the ground-based infrared (IR) echelle spectrograph IRSHELL (\citealt{1991ApJ...376..556L}). Methane is usually identified on IR spectra using the 7.7~$\mu$m ($\sim$1300~cm$^{-1}$) deformation mode absorbance band. \cite{1996A&A...315L.377B} performed the first fit of the methane absorption band at 7.7~$\mu$m with the CH$_4$:H$_2$O ice spectrum and proposed the CH$_4$:CH$_3$OH and CH$_4$:NH$_3$ mixtures as alternatives. Based on the shape and position of the absorbance band, it was concluded that methane is contained within the polar layer of ice mantles on interstellar dust \citep{1998A&A...336..352B}.  Methane was identified in a number of objects in a review of Infrared Space Observatory data \citep{gibb2004interstellar}, where spectra of pure methane and methane mixed with H$_2$O were used. Later, the methane abundance was constrained using \textit{Spitzer} data, where methane was fitted with CH$_4$:H$_2$O mixtures \citep{spitzer_oberg_2008}. Currently, research on methane continues with the recent \textit{James Webb} Space Telescope (JWST) data. The deformation mode of solid methane at 7.7 $\mu$m is described by methane-bearing H$_2$O ices: at 15~K \citep[H$_2$O:CH$_4$ = 10:1\footnote{Mixtures used in fitting procedures in the cited work are shown with this notation.};][]{rocha2024jwst} in IRAS 2A and IRAS 23385+6053, at 13 K \citep[H$_2$O:CH$_4$ = 100:6;][]{chen2024joys+} in B1-c, at 12 K \citep[H$_2$O:CH$_4$ = 10:1;][]{rocha2025ice} in Ced 110 IRS4, for an H$_2$O and CO$_2$ environment at 12 K\footnote{Note that the spectrum that is referred to in the fitting procedure is not listed in the original paper \citep{rocha2017infrared}.} \cite[H$_2$O:CO$_2$:CH$_4$ = 10:1:1;][]{mcclure2023ice} in NIR38 and J110621, using a complex mixture at 10 K \cite[H$_2$O:CH$_3$OH:CO$_2$:CH$_4$ = 0.6:0.7:1:0.1;][]{2022ApJ...941L..13Y} in IRAS 15398--3359, and  H$_2$O:CH$_4$ = 10:1 at $\sim$8 K and CO$_2$:CH$_4$ = 5:1 at  $\sim$27 K \citep{nakibov2025solid} in IRAS 23385+6053. According to these findings, the methane abundance in cold clouds and star-forming regions is 1.6--3.7\% with respect to solid H$_2$O. 

There are three phases of solid methane under vacuum conditions: phase~I is stable above 20.4~K, phase~II is stable below 20.4~K \citep[e.g.,][]{Gerakines_2015}, and the third phase is stable when methane is deposited below 10~K \citep{1962JChPh..36.2223C,PismaZhETF.17.605,emtiaz2019infrared}. Following \cite{emtiaz2019infrared}, we refer to this third phase as the crystalline phase~II*. However, the literature spectra of methane that were typically used to interpret observational data were obtained at temperatures of 10 K and above, and the laboratory data on ice analogs are presented as water-based mixtures for phases I and~II \citep{hudson1999laboratory,1999A&A...350..240E, rocha2017infrared, mifsud2023proton}. Binary mixtures of methane with H$_2$O, CO, CH$_3$OH,  NH$_3$, CO$_2$, and N$_2$, O$_2$ were described for the first time by \cite{1997A&A...317..929B} at 10 K, but the spectral data needed for fitting are not available. 
There are also reflection-mode studies \citep[e.g.,][]{2014PCCP...16.3399K,2014ApJ...789...36M,2017ApJ...845...83F,2018ApJ...852...70B} that characterized methane photochemistry using relevant laboratory ices, but their spectra are not directly comparable to astronomical observations of ices in transmission.

We present laboratory spectra of binary methane-containing ices in the transmission regime that are currently missing in the literature. These mixtures correspond to the environment of the polar layer of ice on the dust grains and are meant to help us interpret JWST data. The ices have astrochemically relevant component ratios: CH$_4$:H$_2$O = 1:10, CH$_4$:CO$_2$ = 1:5 and 1:15, CH$_4$:CH$_3$OH = 1:3, and CH$_4$:NH$_3$ = 1:3. The set of mixtures was obtained at a substrate temperature of 10~K. However, in cold pre-stellar cores, temperatures can reach as low as 6 K, which was confirmed, for instance, by observations of gas and dust in the centers of the isolated cold dense cores L183 and L1544 \citep{2004A&A...417..605P, 2007A&A...467..179P, 2007A&A...470..221C, 2015A&A...574L...5P}. Moreover, the physical modeling of L1544 yielded a dust temperature in the core center equal to 6.5~K~\citep{keto2010dynamics}. The presence of dust colder than 10~K was also confirmed by observations obtained with Submillimetre Common-User Bolometer Array (SCUBA) at 850 $\mu$m \citep{2001ApJ...557..193E, 2015MNRAS.450.1094P}. Therefore, the set was extended with ices grown at 6.7~K with the same composition. The extended set explored the difference between the ices grown at 6.7 K and 10 K substrate temperatures, which correspond to the crystalline II* and crystalline II phases of pure methane. The transmission spectra of methane in binary mixtures grown at 6.7~K and CH$_4$:CH$_3$OH and CH$_4$:NH$_3$ at 10~K are presented here for the first time. The band strengths of methane in various environments are provided as well.

In our previous paper \citep{nakibov2025solid}, we described gaseous and solid methane in IRAS 23385+6053 using some of the spectra presented here. In this paper, we show the utility of our new IR spectra by exploring the spatial distribution of methane in JWST spectra of the variable protostar B335-IRS (hereafter B335). B335 is an isolated nebula at a distance of 90--120 pc with a young class 0 protostar, IRAS19347+0727 \citep[age $\sim$$10^4$ yr;][]{2015A&A...576A.109Y}. It is associated with a protostellar high-velocity Herbig-Haro jet  \citep{1992A&A...256..225R,2007A&A...475..281G}. This source has a high level of accretion and two outflows \citep{2019A&A...631A..64B}, and its luminosity increased by a factor of 5–7 on Modified Julian Date (MJD) 56948 \citep{2023ApJ...943...90E}. This source is a natural laboratory for studying ice chemistry in the early stages of star formation. While the sublimation of complex ices and the rich gas-phase chemistry near the heated core were observed at millimeter wavelengths \citep{2025ApJ...978L...3L}, we focus here on the solid methane features at 7.7 $\mu$m that are located within the outflow cavity wall. This allows us to probe the composition of ices in the cold envelope of a young protostar.

\section{Experimental methods}
All experiments on growing analogs of interstellar ice were carried out on the Ice Spectroscopy Experimental Aggregate (ISEAge), which uses an ultrahigh vacuum setup. A detailed description of our installation is provided in \cite{Ozhiganov_2024}. Briefly, the molecules are deposited on the Ge substrate from two sides via background deposition. A circular area of the window with a diameter of 6 mm (S=0.28 cm$^2$)  is available for the deposition of the ice. The substrate is mounted on a cryogenic finger of a closed-cycle He cryostat, and the whole chamber is continuously evacuated to the base pressure of $2$\texttimes$10^{-10}$~mbar. The operating temperatures are within the 6.7--305 K range with an error of $<$0.1 K. The setup is equipped with a Stanford Research Systems RGA200 quadrupole mass spectrometer and a Thermo Scientific Nicolet iS50 Fourier transform infrared (FTIR) spectrometer operating within 4000--630~cm$^{-1}$ (2.5--15.9~$\mu$m) with 1~cm$^{-1}$ resolution. The points in the collected spectra are spaced at 0.12 cm$^{-1}$ intervals due to the zero filling performed on the interferometer level. The IR spectra are collected in the transmission regime. The FTIR spectrometer and  optical path outside the main chamber are flushed by pure N$_2$. Gaseous species are introduced into the main chamber via two independent leak valves.

We followed the same experimental procedure to grow all ices. The component ratio was controlled during the deposition by means of the quadrupole mass spectrometer (QMS). In all experiments, the two components of a binary mixture were deposited simultaneously. The first component introduced into the main chamber was the pure methane, and the second component was a matrix molecule introduced shortly after the desired ion current of methane was reached. The mixture deposition time started when ion currents for both components reached the desired values. This approach allowed us to correct for the effect of ion interference and overlapping ion currents and to maintain the same ion current for methane in all the experiments. The ion current values were taken from the calibration curves obtained for our setup in a procedure similar to that described by \cite{Slavicinska_et_al._(2023)}. Briefly, the rate at which gaseous molecules are adsorbed on the substrate depends nonlinearly on the ion current and is unique to each molecule. By growing pure ices at different ion currents, we obtain this dependence via an experiment, which gives us more precise control over the component ratio of mixed ices. The rates for the calibration curves were determined by measuring the absorbance on the IR spectrum output using the literature band strength values. Since the depositions were conducted on the Ge substrate cooled to 10 K and 6.7 K, we used the band strength for methane in phase II (corrected for density) from \cite{bouilloud2015bibliographic} and for methane in phase II* from \cite{Gerakines_2015}.

For both sets of experiments, the methane deposition rate was fixed at $5.9\times10^{12}$ cm$^{-2}$ s$^{-1}$, and the second component deposition rate varied based on the desired ratio of the components. The deposition continued for 120 minutes. We obtained a column density of $4.25\times 10^{16}$~cm$^{-2}$ for CH$_4$.  This approach enabled us to compare the methane deformation (1300 cm$^{-1}$) and stretching (3000~cm$^{-1}$) modes in various molecular environments. The IR spectra were recorded every 45 seconds (an average of 32 scans). For the analysis, we used the last recorded deposition spectrum in each experiment to avoid the secondary adsorption effect after closing the leak valve. This effect is present when working with volatile species such as CH$_4$, CO, and N$_2$ in both static conditions and during the warm-up (see Appendix \ref{AppA} for details). To calculate the peak positions and areas, the spectra were baseline-corrected for the overall slope and local thin-film interference artifacts. Repeated measurements (50 iterations) with randomized local baseline adjustments were analyzed to minimize noise effects. The final values were derived from the averaged results.
 
The compounds we used were CH$_4$ (99.999~\%, Ugra-PGS), CO$_2$ (99.9999~\%, Ugra-PGS), NH$_3$ (99.9~\%, UralKrioGaz), and deionized H$_2$O and CH$_3$OH ($\geqslant$99.8~\%, Vekton). Gaseous CO$_2$ and CH$_4$ were introduced directly into the dosing lines from the commercially acquired gas bottles. H$_2$O, NH$_3$, and CH$_3$OH were purified via a freeze-pump-thaw cycle three times before each experiment. Before each deposition, every compound was checked for purity by means of mass spectrometry.    

\section{Results}
\subsection{Binary mixtures with CH$_4$}

\begin{figure*}[!h]
    \sidecaption
    \includegraphics[width=12cm]{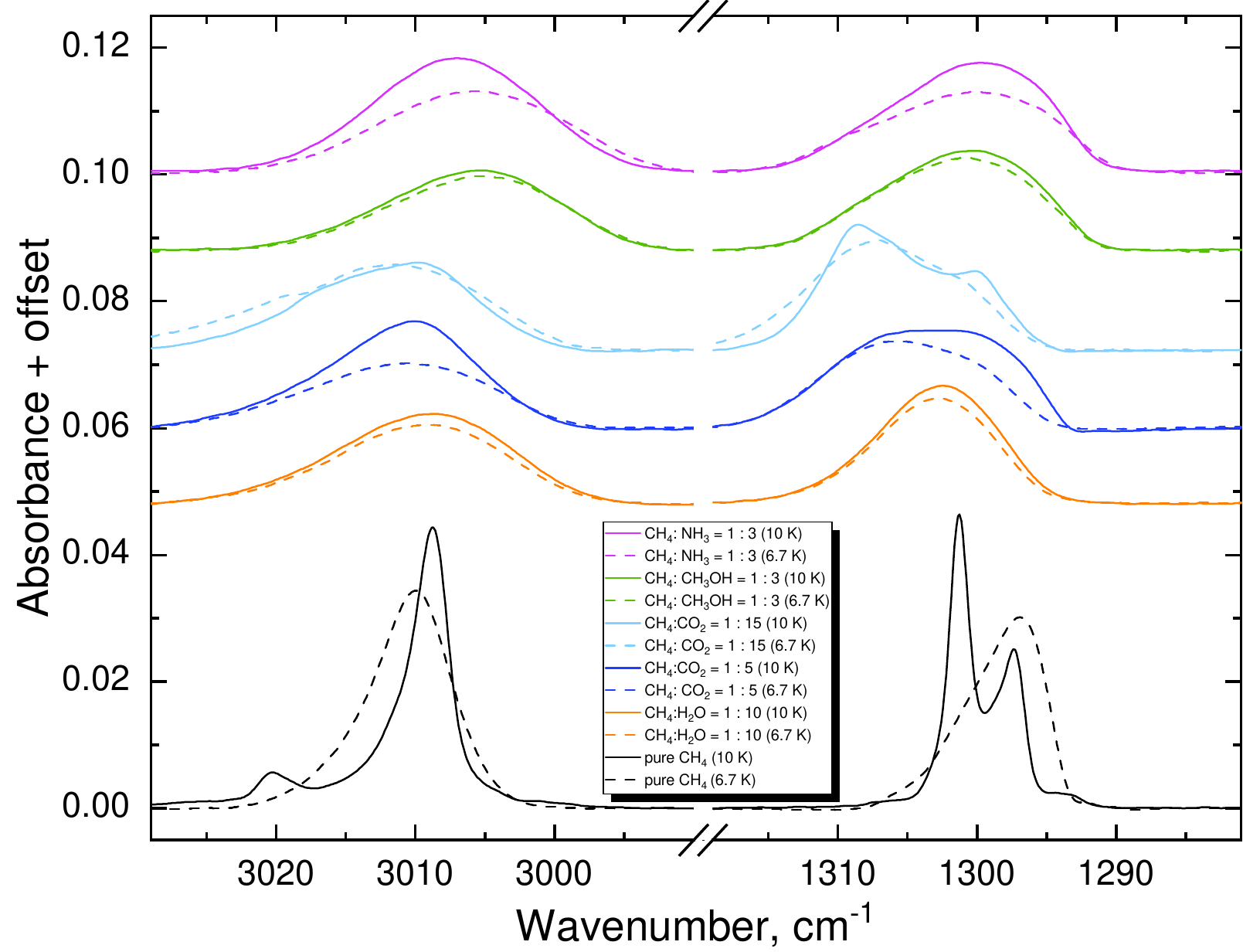}
        \caption{Main vibrational modes of methane at 6.7 K and 10 K: the stretching mode at 3000 cm$^{-1}$ (3.33 $\mu$m, $\nu_3$) and the deformation mode at 1300 cm$^{-1}$ (7.7 $\mu$m, $\nu_4$).\ Black shows pure methane, yellow the CH$_4$:H$_2$O = 1:10 mixture, blue the CH$_4$:CO$_2$ = 1:5 mixture, light blue the CH$_4$:CO$_2$ = 1:15 mixture, green the CH$_4$:CH$_3$OH = 1:3 mixture, and purple the CH$_4$:NH$_3$ = 1:3 mixture. The column density of CH$_4$ in all the IR spectra equals $4.25\times 10^{16}$~cm$^{-2}$.}
            \label{fig1}
\end{figure*}

In this section we present a series of experiments we conducted to obtain the binary ices of methane with molecules that constitute the polar layer of the interstellar ice mantles. We obtained binary mixtures at a temperature of 10~K, with methane molecules embedded in a matrix of H$_2$O, CO$_2$, CH$_3$OH, or NH$_3$. The component ratio was chosen to qualitatively reflect the current astrochemical view on species abundances in interstellar ices in cold dense molecular clouds \citep{boogert2015observations_Ch4,Goto_ea21,mcclure2023ice,Stats}. Additionally, the comparative set was obtained at the minimum stable temperature of the ISEAge setup at 6.7~K. Methane absorption bands in mixtures at 6.7~K spectra differ from both pure methane and 10~K mixture features. We present the spectra of the mixtures for both temperatures in Fig.~\ref{fig1}. Additional experiments concerning pure methane and its transition between phases II* and II are presented in Appendix \ref{AppA}.
The methane absorption bands change shape and position in mixed ices. The component ratio and peak position of the 3000~cm$^{-1}$ stretching and 1300~cm$^{-1}$ deformation mode are presented in Table~\ref{peaks1300dep}. The band strengths for the deformation mode of methane in various mixtures are also included. The methane deformation mode in the CH$_4$:CO$_2$~=~1:5 mixture stands out because of its structure. It displays an unresolved double peak in both the 6.7 K and 10 K experiments, which becomes resolved by heating after 30--40 K \citep{1997A&A...317..929B, nakibov2025solid}. Moreover, this mixture displays the strongest distinction between the two temperatures in both vibrational modes. This is caused by the amorphous and crystalline form of pure CO$_2$, as shown by \cite{2015ApJ...808L..40G}. For our fixed column density of 2.12$\times$10$^{17}$ cm$^{-2}$, pure CO$_2$ fully crystallizes at 10 K and crystallizes only partially at 6.7 K. An addition of methane, however, prevents CO$_2$ from crystallizing at 6.7 K. To further investigate this, we obtained more diluted mixtures: CH$_4$:CO$_2$~=~1:15 at 6.7 K and 10 K. The differences persisted despite the higher CO$_2$ concentration. We therefore report that the peak shape of methane differs significantly in crystalline and amorphous CO$_2$. We also note that the double peak becomes resolved at 10 K in the CH$_4$:CO$_2$~=~1:15 mixture, which is in line with the results of \cite{1993ApJS...86..713H}, who observed the double structure in a CH$_4$:CO$_2$ = 1:20 mixture at 10~K.

Methane mixtures with CH$_3$OH and NH$_3$ show similar features in the 1300 cm$^{-1}$ region and have similar band strength values. The features are redshifted by 2~cm$^{-1}$ from the H$_2$O:CH$_4$ feature. This opens the possibility of singling out methane that is not related to water ice in observational spectra, an idea first introduced by \cite{1996A&A...315L.377B}. This is important, as normally methane is assumed to be formed simultaneously with water and ammonia ice via the hydrogenation of carbon atoms at the onset of the formation of icy grain mantles \citep{2020NatAs...4..781Q,2022ApJ...928...48L,2023ApJ...944..142F}. However, methane can also be formed as a product of the photo-processing or ion-bombardment of methanol-rich ices \citep{2014A&A...561A..73I} in later stages of the ice evolution or via electron-induced radiolysis \citep{2016MNRAS.460..664S}. The presented data could potentially be used to explore the relative importance of the two scenarios of methane formation.

\begin{table*}[!h]
{\small
    \caption{Position of the $\nu_3$ and $\nu_4$ methane modes in mixtures.}
    \label{peaks1300dep}
    \centering
    \begin{tabular}{lcccccc}
    \hline
    \hline
    Mixture & 
    Component ratio & 
    Temperature (K) & 
    Peak position $\nu_3$ (cm$^{-1}$) & 
    Peak position $\nu_4$ (cm$^{-1}$) & 
    Band Strength $\nu_4$ (cm) \\
    \hline
    CH$_4$              & -         & 8     &   3010.0 &    1296.9  & 1.04\texttimes10$^{-17}$\tablefootmark{a} \\
    CH$_4$              & -         & 10    &   3008.8 &    1301.3  & 8.4\texttimes10$^{-18}$\tablefootmark{b} \\  
                        &           &       &   3020.5 &    1297.0                               \\
    \hline
    CH$_4$:H$_2$O     & 1 : 10    & 6.7   &   3008.9 &    1302.7  & 8.8\texttimes10$^{-18}$  \\
    CH$_4$:H$_2$O     & 1 : 10    & 10    &   3008.6 &    1302.5  & 1.0\texttimes10$^{-17}$  \\
    CH$_4$:CO$_2$     & 1 : 5     & 6.7   &   3010.6 &    1305.6  & 9.3\texttimes10$^{-18}$  \\
    CH$_4$:CO$_2$     & 1 : 5     & 10    &   3010.1 &    1301.7  & 1.2\texttimes10$^{-17}$  \\
    CH$_4$:CO$_2$     & 1 : 15    & 6.7   &   3011.7 &    1307.0  & 1.0\texttimes10$^{-17}$  \\
    CH$_4$:CO$_2$     & 1 : 15    & 10    &   3009.8 &    1308.6  & 1.1\texttimes10$^{-17}$  \\
                      &           &       &   3039.9 &    1300.1                                 \\
    CH$_4$:CH$_3$OH   & 1 : 3     & 6.7   &   3005.6 &    1300.8  & 1.0\texttimes10$^{-17}$  \\
    CH$_4$:CH$_3$OH   & 1 : 3     & 10    &   3005.2 &    1300.3  & 1.1\texttimes10$^{-17}$  \\
    CH$_4$:NH$_3$     & 1 : 3     & 6.7   &   3005.6 &    1300.2  & 1.0\texttimes10$^{-17}$  \\
    CH$_4$:NH$_3$     & 1 : 3     & 10    &   3007.0 &    1299.7  & 1.2\texttimes10$^{-17}$  \\
    \hline
    \end{tabular}
    \tablefoot{
    \tablefoottext{a}{The band strength is taken from \cite{Gerakines_2015}.} 
    \tablefoottext{b}{The band strength is taken from \cite{bouilloud2015bibliographic}.}
    }
}
\end{table*}

\subsection{B335 methane survey}

To test the usability of the obtained spectra, we fitted the 7.7~$\mu$m methane absorption band of the Class 0 protostar B335. JWST Mid-IR Instrument (MIRI) Medium Resolution Spectrograph (MRS) observations toward B335 are provided within the Investigating Protostellar Accretion proposal \citep[PID 1802, P.I. T. Megeath;][]{2021jwst.prop.1802M}. Level 3 observation data were taken from the Mikulski Archive for Space Telescopes (MAST) database (DOI: 10.17909/yr06-tf19), and we further scaled the CH2-SHORT channel data to match the CH1-LONG signal level. MIRI MRS observations included three levels of the JWST calibration pipeline with version 1.19.1 and Calibration References Data System (CRDS) context jwst\_1413.pmap (CRDS\_VER = '12.1.11'), which is described in  \cite{Greenfield2016}, \cite{Bushouse2024}, and \cite{Gelder2024}. Unfortunately, the overlap region coincides with the OCN$^-$ absorption band, and we excluded this species from further analysis.

\begin{figure}[!h]
    \centering
    \includegraphics[width=\hsize]{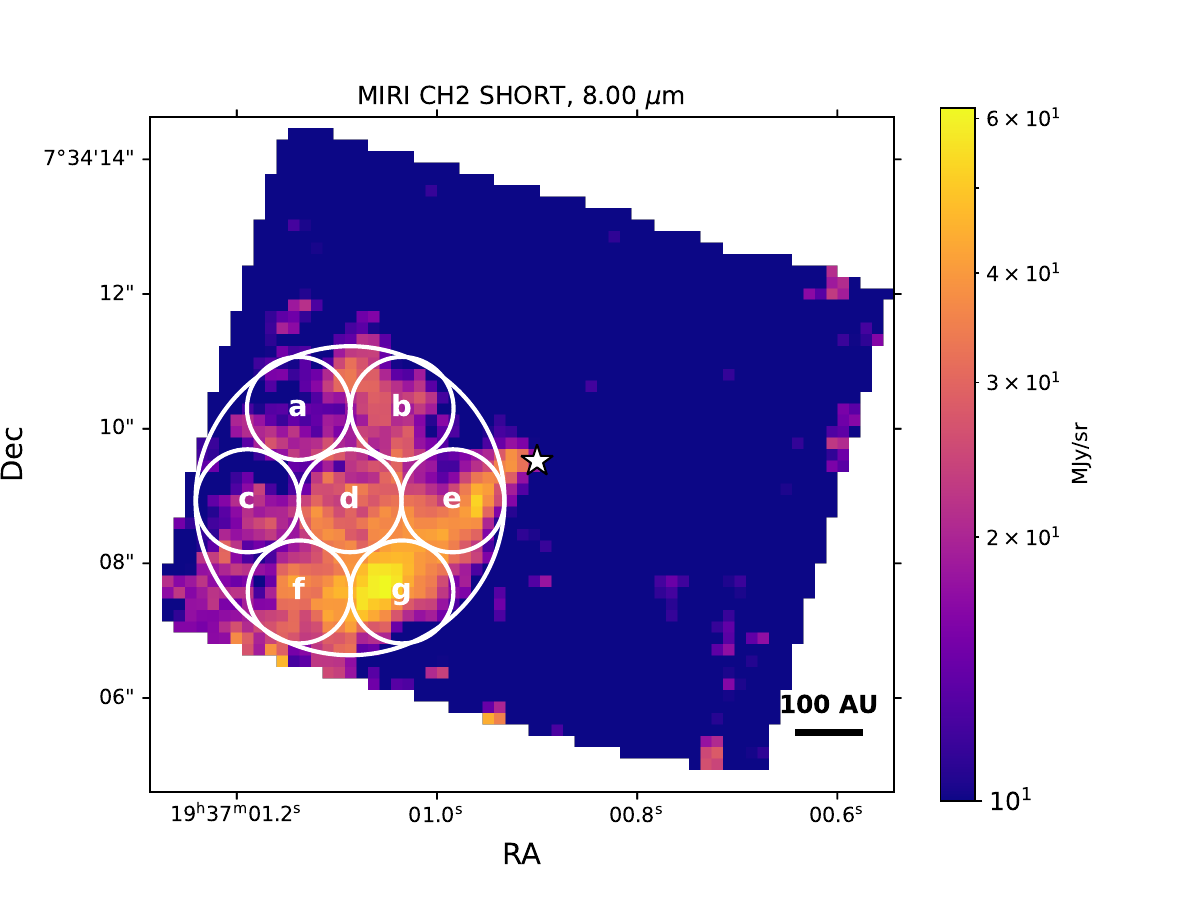}
    \caption{B335 intensity map at 8 $\mu$m using the JWST MIRI-MRS spectrograph data. The small circles denote the 1.53$^{\prime\prime}$ apertures, and the large circle denotes the 4.59$^{\prime\prime}$ aperture. The center coordinates of the chosen apertures are listed in Table \ref{apertures}.}
    \label{fig2}
\end{figure}

The object B335 is viewed edge-on and has two prominent outflows \citep[e.g.,][]{2019A&A...631A..64B}, the western of which displays strong continuum emission around 7.7 $\mu$m. Based on the intensity map at 8 $\mu$m (Fig. \ref{fig2}), we selected multiple apertures that covered bright pixels. This approach maximizes the signal-to-noise ratio, which facilitates our analysis of the finer spectral details. This is the first study that explores the spatial distribution on MIRI-MRS JWST data for a molecule with a low abundance relative to H$_2$O. We note that a spatial distribution study was carried out recently for solid H$_2$O, CO, and CO$_2$ on NIRCam data \citep{2025NatAs...9..883S}. The spectra were extracted from the cylindrical apertures: seven 1.53$^{\prime\prime}$ apertures were used to assess the spatial distribution of methane, and a 4.59$^{\prime\prime}$ aperture was used to ensure consistency between fits in smaller apertures. The center positions for the chosen apertures are listed in Table \ref{apertures}.

All spectra were processed in the same way. After baseline subtraction, a linear combination of methane laboratory mixtures was fitted to the observational spectrum. The linear scale coefficients were obtained by minimizing $\chi^2$, given as
\begin{equation}
    \chi^2=\sum^{N_{obs}}_{i=1}\frac{(\tau_{obs}^i-\tau_{model}^i)^2}{\sigma_i^2},
\end{equation}
where $\tau_{obs}$ and $\tau_{model}$ are observational and modeled optical depths, respectively, and $\sigma_i$ is the point-wise standard deviation of the observational spectrum. The standard deviation was obtained from the residuals between the original optical depth spectrum and a smoothed curve obtained with locally weighted scatterplot smoothing \citep[][]{Cleveland_1979} with a 24-point window. The column densities of H$_2$O and CO$_2$ were estimated based on the 6 $\mu$m and 15.2 $\mu$m bands, respectively. 

\begin{figure*}[h]
    \centering
    \includegraphics[width=\hsize]{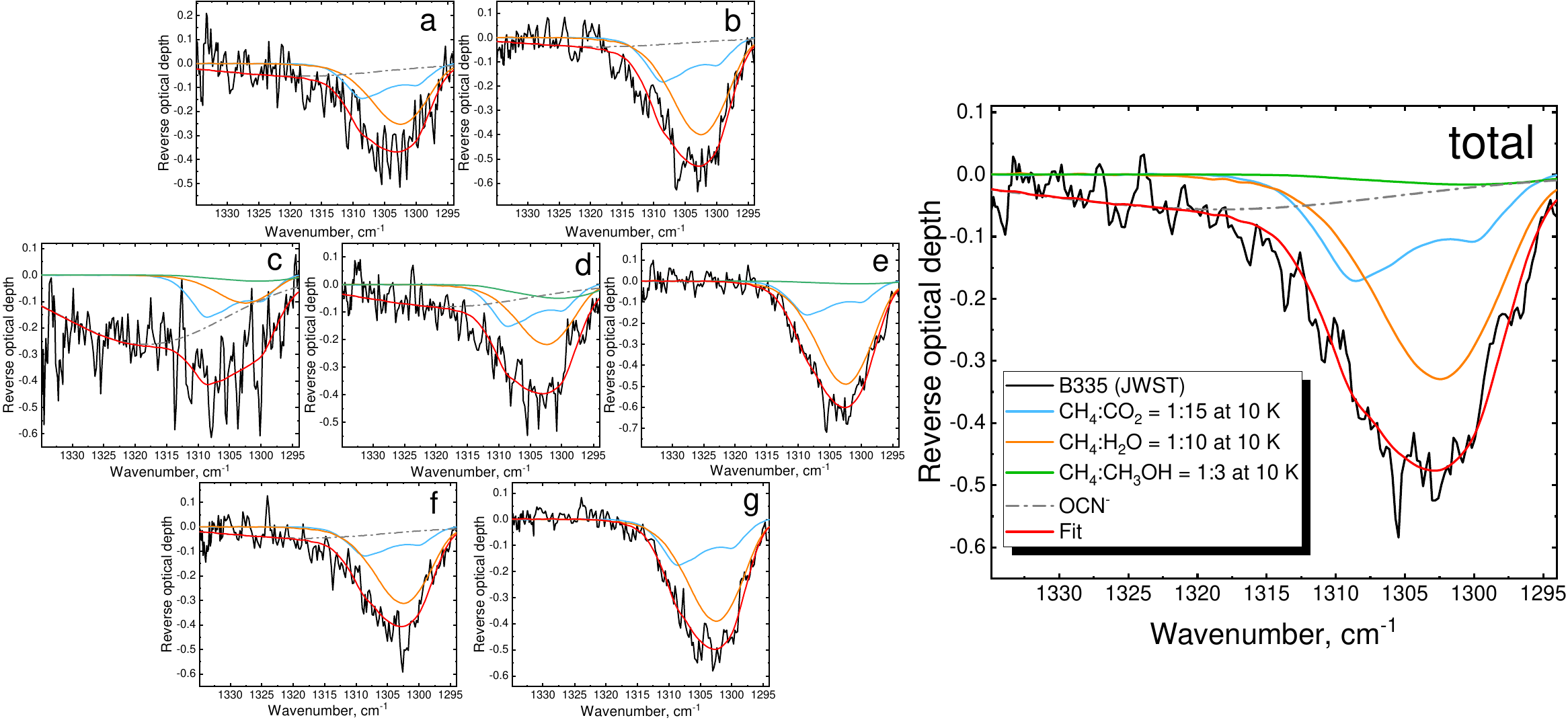}
    \caption{Observational data fitted with ISEAge laboratory mixtures (solid lines). We show the CH$_4$:H$_2$O = 1:10 mixture at 10~K (orange), CH$_4$:CO$_2$ = 1:15 at 10~K (blue), CH$_4$:CH$_3$OH = 1:3 at 10 K (green), and OCN$^-$ (gray). The final fit is presented in red.}
    \label{fig3}
\end{figure*}

The fitted spectra are shown in Fig. \ref{fig3}. The $a$-$g$ observational spectra correspond to the $a$-$g$ apertures from Fig.~\ref{fig2}. The fitting results show that about 70\% and 30\% of the methane budget in B335 are best described by methane in H$_2$O and methane in CO$_2$ environments, respectively (see Table \ref{apertures}). A minor fraction of the methane is found in the CH$_3$OH environment, but the detection at this level may be sensitive to baseline selection, for example. Spectrum $c$ was excluded from the analysis because the signal-to-noise ratio was low. The 1.53$^{\prime\prime}$ apertures display variations in ice composition, with CH$_4$ in CO$_2$ to CH$_4$ in H$_2$O relative abundances decreasing toward the protostar. The column density toward the core also increases overall, which is likely explained by an increase in density toward the center. 

The presence of CH$_4$ in the CO$_2$ matrix provides some insights into ice formation in early stages of the protostar. An interesting correlation is observed between CH$_4$ and CO$_2$. The Spearman $\rho$ and p values \citep{Spearman} for the CO$_2$ and CH$_4$ column densities ($\rho=0.9$, $p$=0.007) and abundances ($\rho=0.9$, $p$=0.003) indicate a strong statistically significant positive correlation between these species. The correlations between CH$_4$ and CO$_2$ are known from the \textit{Spitzer} methane survey \citep{spitzer_oberg_2008}. The modeling presented in \cite{2011ApJ...735...15G} shows that under certain conditions, methane and CO$_2$ can form concurrently, occupying the innermost layers of ice mantles. \cite{2024ApJ...960...22F} explored the reactivity of the $^3$C-OH$_2$ complex with various species, revealing barrierless reactions with atomic O and H. These reactions lead to the production of CH$_2$ and CO, which are the direct precursors of CH$_4$ and CO$_2$, respectively. While the parallel formation of methane and CO$_2$ has not been largely explored, these results can be interpreted as evidence that these processes occur. 

\begin{table*}[h]
{\small
\caption{Aperture parameters and values derived from the fit.}
\label{apertures}
\centering
\begin{tabular}{ccccccccc}
\hline
\hline
    Aperture & 
    Coordinates &
    $N$(CH$_4$) & 
    CH$_4$:H$_2$O & 
    CH$_4$:CO$_2$ & 
    CH$_4$:CH$_3$OH &
    $N$(CO$_2$) &
    $N$(H$_2$O) &
    $N$(CH$_4$)/$N$(H$_2$O) \\
     & 
     &
    [10$^{17}$ cm$^{-2}$] &
    [\%] & 
    [\%] &  
    [\%] &
    [10$^{18}$ cm$^{-2}$] &
    [10$^{19}$ cm$^{-2}$] &
    [\%] \\
    \hline
    total   & 19:37:01.0868 & 5.0 & 64 & 32 & 4 & 2.8 & 1.2 & 4.3 \\
            & +07:34:08.9309 &&&&&&& \\
    a       & 19:37:01.1384 & 3.8 & 65 & 35 &  -  & 2.5 & 1.0 & 3.8 \\
            & +07:34:10.3022 &&&&&&& \\
    b       & 19:37:01.0348 & 5.6 & 70 & 30 &  -  & 3.1 & 1.2 & 4.8 \\
            & +07:34:10.3022 &&&&&&& \\
    c       & 19:37:01.1892 & 2.8 & 37 & 53 & 10 & 2.4 & 1.1 & 2.5 \\
            & +07:34:08.9298 &&&&&&& \\
    d       & 19:37:01.0868 & 4.1 & 52 & 34 & 14 & 2.2 & 1.2 & 3.4 \\
            & +07:34:08.9309 &&&&&&& \\
    e       & 19:37:00.9841 & 6.4 & 75 & 23 & 2 & 3.9 & 1.9 & 3.4 \\
            & +07:34:08.9298 &&&&&&& \\
    f       & 19:37:01.1375 & 4.1 & 74 & 26 &  -  & 2.4 & 1.3 & 3.1 \\
            & +07:34:07.5764 &&&&&&& \\
    g       & 19:37:01.0350 & 5.4 & 70 & 30 &  -  & 3.0 & 1.8 & 3.1 \\
            & +07:34:07.5764 &&&&&&& \\
    \hline
\end{tabular}
    \tablefoot{The apertures are listed following the notation in Fig. \ref{fig2}. The coordinates are given for the aperture centers in the form of RA and Dec. $N$(X) denotes the column density of species X. The columns denoted CH$_4$:X (X=H$_2$O, CO$_2$, and CH$_3$OH) show the contribution of methane in a CH$_4$:X mixture to the total CH$_4$ column density listed in Col. 3. $N$(CH$_4$)/$N$(H$_2$O) is the CH$_4$ abundance relative to solid H$_2$O.
    }
}
\end{table*}

\section{Astrochemical implications and conclusions} 

We have presented the spectra of pure methane and binary mixtures with methane for 6.7 K and 10 K deposition temperatures: CH$_4$:H$_2$O~=~1:10, CH$_4$:CO$_2$~=~1:5, CH$_4$:CO$_2$~=~1:15, CH$_4$:CH$_3$OH~=~1:3, and CH$_4$:NH$_3$~=~1:3. All spectra of methane in astrochemically relevant mixture ratios are presented for the first time at 6.7 K. The new band strength values and spectra obtained at a temperature below 9 K revealed the astrochemical implications listed below. 

1. The band strength of pure methane at 10 K is currently mainly used for column density estimations in observational data \citep[e.g.,][]{mcclure2023ice,rocha2024jwst,rocha2025ice}. However, solid interstellar methane is found in mixtures with other icy species rather than in pure form. We also note that interstellar ices differ from laboratory analogs both in how they are produced and in how they are processed. Thus, while the band strengths in Table~\ref{peaks1300dep} provide a better basis for interpreting observations, they are still approximate. Nevertheless, our new data invite a discussion on recent literature estimates. Since the $\nu_4$ methane band strengths in mixtures are $\approx$20~\% higher on average than the band strength of pure methane at 10~K, the column densities of interstellar methane may be overestimated by this value by~\citet{mcclure2023ice}, for instance. Interestingly, this partially explains the underestimation of the solid methane abundance by astrochemical models toward the background stars NIR38 and J110621 reported in~\citet{Stats}. 

2. We have presented the the first publicly available digital spectral library of methane absorbance spectra in a non-H$_2$O environment, providing the possibility to identify non-H$_2$O methane environments in newly obtained JWST spectra. 
Solid methane is typically considered to reside in a polar water-rich environment because it forms early along with water ice \citep[see, e.g.,][]{1992ApJS...82..167H,2023A&A...675A.165C,Borshcheva_ea25}. This is further supported by successful fits of methane features in JWST spectra of interstellar ices with CH$_4$:H$_2$O mixtures \citep[e.g.,][]{chen2024joys+,rocha2024jwst,rocha2025ice}. However, there is possible evidence that it is present in mixtures with other molecules. For example, \cite{1996A&A...315L.377B} and \cite{nakibov2025solid} show that the observed shapes of the 7.7~$\mu$m methane feature can be partially interpreted with non-H$_2$O methane-containing mixtures. Additionally, icy methane mixed with apolar species such as N$_2$ and CO is found on icy bodies in the outer Solar System \citep[e.g.,][]{doi:10.1021/acs.jpca.2c00287}. The obtained spectra of ices of astrochemical ratios are published on Zenodo (DOI: \href{https://doi.org/10.5281/zenodo.15350566}{10.5281/zenodo.15350566}). Our comparison of spectra obtained at 6.7 K and 10 K revealed temperature-induced changes in shape and peak position, although we find it unlikely that these differences are prominent enough to be discerned within the observational data. However, it is known that the central regions of pre-stellar cores are well shielded from external heating radiation. Consequently, as shown both observationally \citep{2004A&A...417..605P, 2007A&A...467..179P, 2007A&A...470..221C, 2015A&A...574L...5P} and theoretically \citep[][and references therein]{2005ApJ...635.1151K,keto2010dynamics}, the dust temperature there can reach 6~K. According to recent modeling by \citet{Borshcheva_ea25}, approximately 80\% of solid methane in the pre-stellar core L1544 resides on grains with temperatures below 9~K. Thus, to accurately interpret future observations of ices on a line of sight toward L1544, laboratory spectra and band strengths of mixtures obtained at both 6.7~K and 10~K might be needed. The presented spectra  can be directly compared to observational data because the $\nu_4$ deformation mode of methane is unlikely to be affected by grain size; it mostly depends on temperature and the molecular environment \citep{1997A&A...317..929B}.

3. We have presented the first spatial survey performed on JWST MIRI-MRS data and explored methane in B335. The results show that about 30\% of solid methane is found in a CO$_2$ environment and not in the H$_2$O matrix. The consistency between the findings from full and smaller apertures proves that the spectra we have presented can be used to analyze JWST data. The observed ice composition gradient further demonstrates that JWST provides enough sensitivity for such studies, which should motivate further research.

\section*{Data availability}
The laboratory IR absorbance spectra in the transmission regime in the 2.5--15.9 $\mu$m (4000--630 cm$^{-1}$) region for CH$_4$ in different environments at 6.7 K and 10 K can be found on Zenodo via the following link: \href{https://zenodo.org/records/15350566}{https://zenodo.org/records/15350566}.

\begin{acknowledgements}
     The authors thank Prof. Dr. Paola Caselli for the discussion on the dust temperature in prestellar cores, and Ms. Katerina Borshcheva for providing the information on the distribution of methane ice in the model of L1544. We thank the anonymous reviewer for their comments that helped us to improve the manuscript and A\&A Language Editor Astrid Peter for corrections in the English text. This research work is funded by the Russian Science Foundation via 23-12-00315 agreement.
\end{acknowledgements}

\bibliographystyle{aa} 
\bibliography{Karteyeva} 

\begin{appendix}

\onecolumn
\section{Pure CH$_4$}\label{AppA}

There are three phases of solid methane under vacuum conditions. The crystalline phase II is stable below 20.4~K and has the \textit{fcc} lattice with eight sub-lattices: six of which have site symmetry D$_{2d}$ and two of which have site symmetry O$_{h}$ \citep{10.1063/1.433879}. It displays a double peak structure of both $\nu_3$ (3009 cm$^{-1}$ and 3021 cm$^{-1}$) and $\nu_4$ (1301 cm$^{-1}$ and 1298 cm$^{-1}$) features. This phase reversibly transitions into phase I above the first transition temperature of 20.4~K and has \textit{fcc} lattice with a single freely rotating molecule. The $\nu_3$ and $\nu_4$ features are positioned at 3010 cm$^{-1}$ and 1299 cm$^{-1}$, respectively. Phase II* is observed if methane ice is obtained by vapor deposition below 10 K. It displays a degree of orientational disorder between phase I and phase II \citep{emtiaz2019infrared}. In this phase the absorption bands of methane are positioned near 3010 cm$^{-1}$ and 1297 cm$^{-1}$ \citep{Gerakines_2015}. Phase II* cannot be obtained by cooling methane ice from the triple point or from phase II ice. On warming phase II* ice undergoes irreversible transition to phase II. The second transition temperature depends on experimental conditions due to metastability and is observed below 10~K \citep{1962JChPh..36.2223C,PismaZhETF.17.605,emtiaz2019infrared}.

Despite the extensive research carried out on methane solid phases in the IR at low temperatures in recent decades \citep{10.1063/1.433879,1993ApJS...86..713H, 1997A&A...317..929B,1998A&A...333.1025M, Gerakines_2015, 2015PCCP...1712545H, emtiaz2019infrared, gerakines2020modified}, there is still an ambiguity between published studies. In several works  \citep{1993ApJS...86..713H,1996A&A...312..289G,1998A&A...333.1025M,2003CPL...378..178A,2011A&A...531A.160D} methane is claimed to be in phase II, but the presented spectral features differ significantly from the theoretically derived shape \citep{10.1063/1.433879}. \cite{Gerakines_2015} addressed this discrepancy by providing spectra of `amorphous methane', suggesting that the literature spectra in the works above display partial amorphous-to-crystalline transition, which was caused by the high deposition rates and experimental constraints of the setups used.

\begin{figure}[!ht]
    \sidecaption
    \includegraphics[width=12cm]{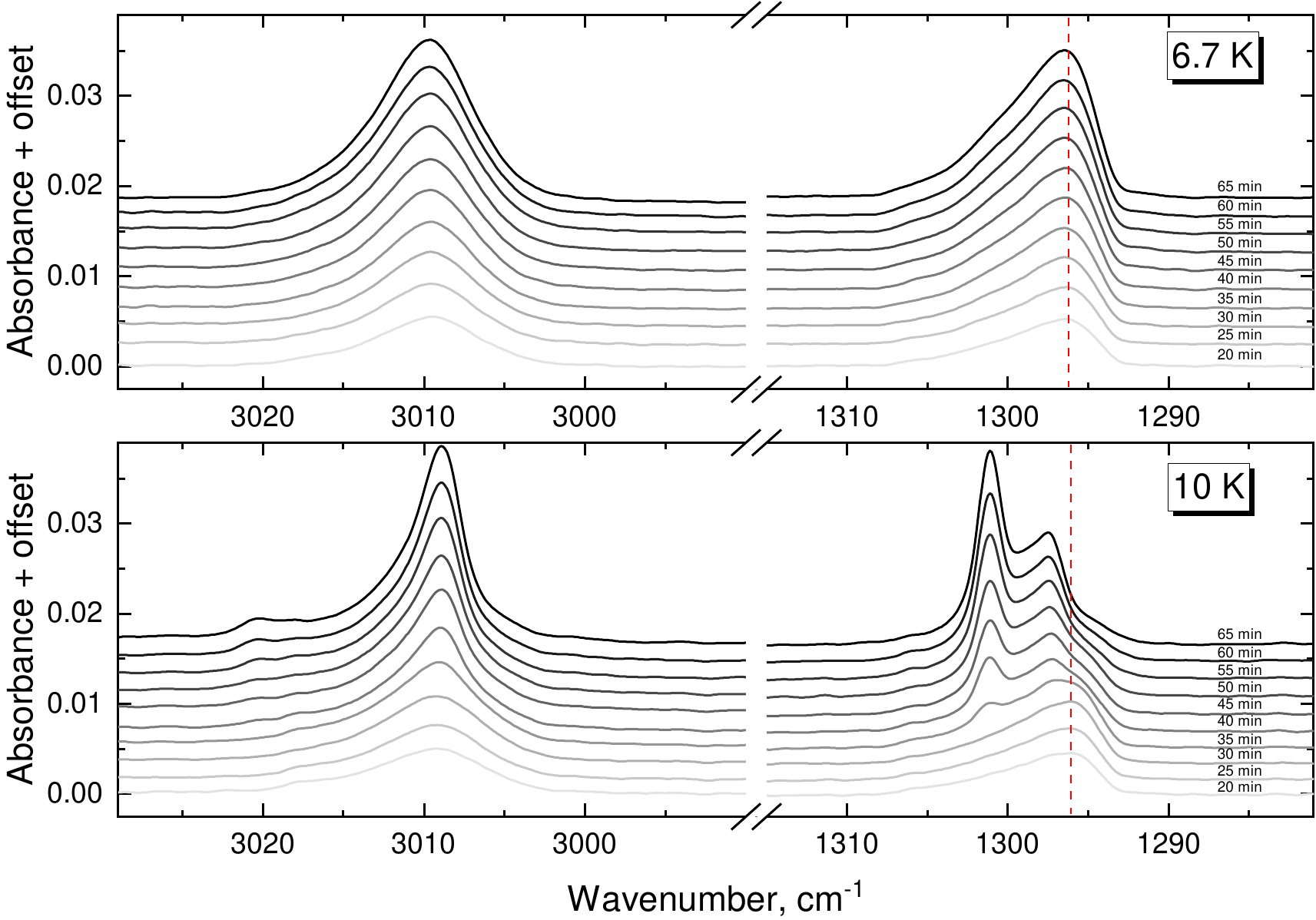}
    \caption{Top panel: Selected IR spectra of pure methane deposition from 20th to 65th minutes with 6.7 K temperature (column density (0.7 to 2.3)$\times10^{16}$~cm$^{-2}$). Bottom panel: IR spectra of pure methane deposition from 20th to 65th minutes with 10 K temperature (column density (0.7--2.3)$\times10^{16}$~cm$^{-2}$). The dashed red line shows the peak position of methane in phase II*.}
    \label{figA}
\end{figure}

\cite{Gerakines_2015} report that the majority of their spectra were obtained in the 8--12 K range, but the spectra are referred to as obtained `at 10 K' for simplicity throughout the text. Thus, it is not clear at what temperature exactly the published experiments were carried out, a point that also applies to \cite{2015PCCP...1712545H}. This ambiguity is significant, because the 8--12~K range includes published methane second transition temperatures \citep{1962JChPh..36.2223C,PismaZhETF.17.605}. We attempted to obtain the spectrum in their Fig. 1f at precisely 10 K, but failed even using a lower deposition rate than stated in their paper. Therefore, prior to obtaining the set of methane-bearing mixtures at 10~K we decided to investigate the transition between phase II* and phase II. Under the deposition rate and duration of experiments described in the main text methane retained phase II* structure when deposited at 6.7 K throughout the whole deposition time. In contrast, for deposition at 10 K, the double peak structure characteristic of phase II started to form once the ice reached certain column density; see Fig. \ref{figA}. At a deposition rate of 5.9$\times10^{12}$ cm$^{-2}$ s$^{-1}$ the transition started after 35 minutes, corresponding to a column density of 1.2$\times$10$^{16}$ cm$^{-2}$. We also conducted depositions at rates twice as low and twice as high: 2.95$\times10^{12}$ cm$^{-2}$ s$^{-1}$ and 1.18$\times10^{13}$ cm$^{-2}$ s$^{-1}$. In all cases, on further deposition the double peak evolved through intermediate shapes, similar to those shown by \cite{Gerakines_2015}, before fully converting to the phase II structure. The change in peak structure occurs at a critical value of 1.2$\times$10$^{16}$ cm$^{-2}$. This equals approximately to 6 monolayers of pure methane ice on each side of the Ge substrate, given that a single monolayer column density is 1$\times$10$^{15}$ cm$^{-2}$.

From our experiments we conclude that phase II* is stable at 6.7 K. At 10~K phase II* is only observed in very thin methane films. In our experiments on Ge substrate, the maximum ice column density at which phase II* is still observed at 10~K is 1.2$\times$10$^{16}$~cm$^{-2}$, independent of the deposition rate. Further ice growth at 10 K leads to the transition of phase II* to phase II through intermediate spectral shapes. While these intermediate spectra are similar to those mentioned in \cite{Gerakines_2015} we refrain from a more direct comparison. We expect that similar results can be obtained under minor differences in experimental conditions of other setups, which is expected given the environment-sensitive nature of metastable phase transition. A definitive answer would come from X-ray diffractometry study as performed in \cite{2016JPSJ...85l4602M,2022JChPh.156c4503M}, which showed that other molecular ices that are `conventionally amorphous' for astrochemistry can exhibit positionally ordered but orientationally disordered structure (CO or N$_2$), while others (CO$_2$, CS$_2$) are indeed amorphous. 

We also note an instrumental effect that concerns both temperature programmed desorption (TPD) and static temperature experiments with volatile species. Here, we interpret the cases transparently reported by \cite{2004MNRAS.354.1133C} and \cite{doi:10.1021/acs.jpca.5c03186}. Methane temporarily adsorbs on warm surfaces (some cryostat parts, radiation shield, etc.) during the deposition and continuously releases even when the leak valves are closed, as in \cite{doi:10.1021/acs.jpca.5c03186}. The evacuation therefore is governed by the amount of temporarily stored methane rather than by the turbomolecular pump's efficiency, resulting in a `tail' in the QMS signal after deposition. During TPD warm surfaces release stored material early, polluting the QMS signal, as in \cite{2004MNRAS.354.1133C}. Released gas can get re-adsorbed by the cold substrate. An example of that can be found in the cyclic experiment in \citet[see their Fig. 1, right column]{Gerakines_2015}. We note that the integrated intensity increased after the first heating-holding-cooling cycle, indicating an increase in column density. The second cycle did not cause an increase because the stored methane was depleted on the first heating. This increase did not affect the results of their paper.

\end{appendix}
\end{document}